\documentclass[aps,reprint,superscriptaddress, showpacs]{revtex4-1}
\usepackage{amsmath}
\usepackage{amssymb}
\usepackage{bm}
\usepackage{graphicx}
\graphicspath{{./figures/}}

\newcommand{\mvec}[1]{\ensuremath{\mathbf{#1}}}
\newcommand{\unit}[1]{\ensuremath{\hat{#1}}}

\newcommand{\Heff}{\ensuremath{\mvec{H}_\mathrm{eff}}}
\def\mean#1{\langle #1 \rangle}

\usepackage[svgnames]{xcolor}

\begin{document}
\title{Driving magnetic skyrmions with microwave fields}

\begin{abstract}
We show theoretically by numerically solving the Landau-Lifshitz-Gilbert equation with a 
classical spin model on a two-dimensional system that both magnetic skyrmions and skyrmion lattices can be moved with 
microwave magnetic fields.
The mechanism is enabled by breaking the axial symmetry 
of the skyrmion, for example through application of a static in-plane external field. The net velocity of the skyrmion 
depends on the frequency and amplitude of the microwave fields as well as the strength of the in-plane field. 
The maximum velocity is found where the frequency of the microwave coincides with the resonance frequency of 
the breathing mode of the skyrmions.
\end{abstract}


\author{Weiwei Wang}
\affiliation{Engineering and the Environment, University of Southampton, Southampton, SO17 1BJ, United Kingdom}
\author{Marijan Beg}
\affiliation{Engineering and the Environment, University of Southampton, Southampton, SO17 1BJ, United Kingdom}
\author{Bin Zhang}
\affiliation{Institut f\"{u}r Experimentalphysik, Freie Universit\"{a}t Berlin, Arnimallee 14, 14195 Berlin, Germany}
\author{Wolfgang Kuch}
\affiliation{Institut f\"{u}r Experimentalphysik, Freie Universit\"{a}t Berlin, Arnimallee 14, 14195 Berlin, Germany}
\author{Hans Fangohr}
\email{fangohr@soton.ac.uk}
\affiliation{Engineering and the Environment, University of Southampton, Southampton, SO17 1BJ, United Kingdom}

\maketitle

Skyrmions, topologically stable magnetization textures with particle-like properties, have recently attracted
great attention~\cite{Rossler2006, Muhlbauer2009, Nagaosa2013, Fert2013} due to their potential use in future 
spintronic devices~\cite{Romming2013}. 
The manipulation of skyrmions is of great importance and interest: skyrmions can be driven using spin-polarized 
current~\cite{Zang2011, Lin2013, Iwasaki2013, Iwasaki2013b}, magnetic or electric field gradients~\cite{Everschor2012, Liu2013}, 
temperature gradients~\cite{Kong2013,Lin2014,Mochizuki2014} and magnons~\cite{Iwasaki2014, Schutte2014, Oh2015}.
Microwaves, on the other hand, have been broadly used in studying various magnetic phenomena, such as the 
ferromagnetic resonance (FMR) and spin wave excitations in skyrmion crystals~\cite{Mochizuki2012, Onose2012, Okamura2013, Schwarze2015}.
However, the possibility of creating translational motion of skyrmions has not been explored in these experiments \cite{Onose2012, Okamura2013, Schwarze2015}. 
In this Rapid Communication, we show that both a single skyrmion and a skyrmion lattice can be moved by microwave fields if 
the axial symmetry of skyrmions is slightly broken by a static in-plane external field.

We employ skyrmions stabilized by the Dzyaloshinskii-Moriya Interaction (DMI)~\cite{Dzyaloshinskii1958, Moriya1960}.
More precisely, the bulk DMI is considered so that a chiral skyrmion (vortex-like) rather than a hedgehog (radial) skyrmion 
configuration emerges~\cite{Muhlbauer2009, Neubauer2009, Zhou2014}.
We start with a classical Heisenberg model on a two-dimensional regular square lattice with nearest-neighbor
symmetric exchange interaction, the bulk-type DMI, and the Zeeman field~\cite{Yi2009, Mochizuki2012, Kong2013}.
In addition, a time-dependent magnetic field $\mvec{h}(t)$ is applied in the $+z$-direction.
Accordingly, the system's Hamiltonian can be written as
\begin{eqnarray}\label{eq_ham}
\mathcal{H}=&-J \sum_{\mean{i,j}} \mvec{m}_i\cdot \mvec{m}_{j} + \sum_{\mean{i,j}} \mvec{D}_{ij}\cdot [\mvec{m}_i \times \mvec{m}_j] \nonumber\\
&- \sum_{i} |\bm \mu_i|  (\mvec{H}+\mvec{h}(t)) \cdot  \mvec{m}_i,
\end{eqnarray}
where $\mean{i,j}$ represents a unique pair of lattice sites $i$ and $j$, $\mvec{m}_i$ is the
unit vector of the magnetic moment $\bm \mu_i = -\hbar \gamma \mvec{S}_i$ with $\mvec{S}_i$ being
the atomic spin and $\gamma(>0)$ the gyromagnetic ratio, and $J$ is the symmetric exchange energy constant.
In the case of bulk DMI, the DMI vector $\mvec{D}_{ij}$ can be written as $\mvec{D}_{ij} = D \hat{\mvec{r}}_{ij}$,
where $D$ is the DMI constant and $\mvec{\hat{r}}_{ij}$ is the unit vector between $\mvec{S}_i$ and $\mvec{S}_j$.
We use the DMI value with $D/J=0.18$, which results in the spiral period $\lambda \sim 2 \pi J a/D \sim 25\,\text{nm}$
for a typical lattice constant $a=0.5$ nm~\cite{Iwasaki2013}.
We consider two nonzero components for the static external field $\mvec{H}$: an in-plane component $H_y$ and a perpendicular
component $H_z$, i.e., $\mvec{H}=(0, H_y, H_z)$.
A nonzero $H_z$ is essential for stabilizing the skyrmion crystal~\cite{Mochizuki2012}.

\begin{figure}[tbhp]
\begin{center}
\includegraphics[scale=0.70]{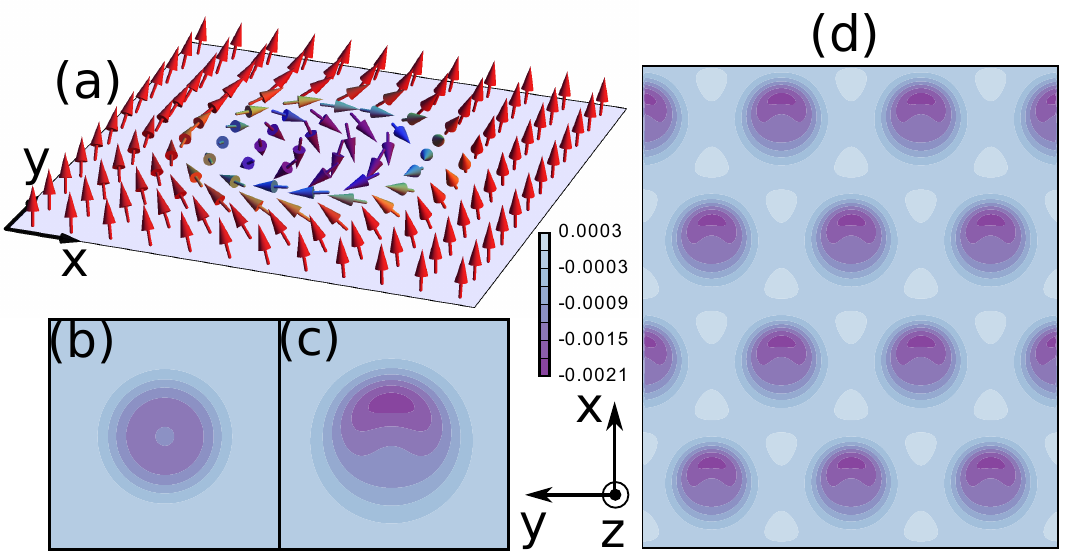}
\caption{(a) Skyrmion configuration in the presence of an in-plane field $H_y=0.006$ with $D=0.18$, $J=1$ and $H_z=0.02$. 
(b) The symmetric topological charge density distribution of a skyrmion when $H_y=0$.
(c) The corresponding topological charge density for the skyrmion shown in (a) when $H_y > 0$.
(d) The skyrmion lattice with 12 skyrmions in a sample of size $N=174 \times 150$ sites, with an in-plane field $H_y=0.004$.}
\label{fig_skx}
\end{center}
\end{figure}

The spin dynamics at lattice site $i$ is governed by the Landau-Lifshitz-Gilbert (LLG) equation,
\begin{equation}\label{eq_llg}
\frac{\partial \mvec{m}_i}{\partial t} = - \gamma \mvec{m}_i \times \Heff + \alpha \mvec{m}_i \times  \frac{\partial \mvec{m}_i}{\partial t}
\end{equation}
where $\alpha$ is the Gilbert damping and $\Heff$ is the effective
field that is computed as $\Heff = -(1/|\bm \mu_i|) {\partial \mathcal{H}}/{\partial \mvec{m}_i}$.
The Hamiltonian~(\ref{eq_ham}) associated with the LLG equation~(\ref{eq_llg}) can be understood as a finite-difference-based micromagnetic model.
Therefore, our simulation results are reproducible by setting the saturation magnetization $M_s=\hbar\gamma S/a^3$, exchange constant $A=J/2a$ and 
DMI constant for continuum form $D_a = -D/a^2$ (corresponding to the energy density $\varepsilon_{\mathrm{dmi}}=D_a \mvec{m}\cdot (\nabla \times \mvec{m})$) 
in micromagnetic simulation packages such as OOMMF~\cite{Porter1999}.
We have carried out simulations with and without dipolar interactions, and the results are qualitatively the same. 
We report results without dipolar interactions for clarity of the model assumptions.

\begin{figure}[tbhp]
\begin{center}
\includegraphics[scale=0.65]{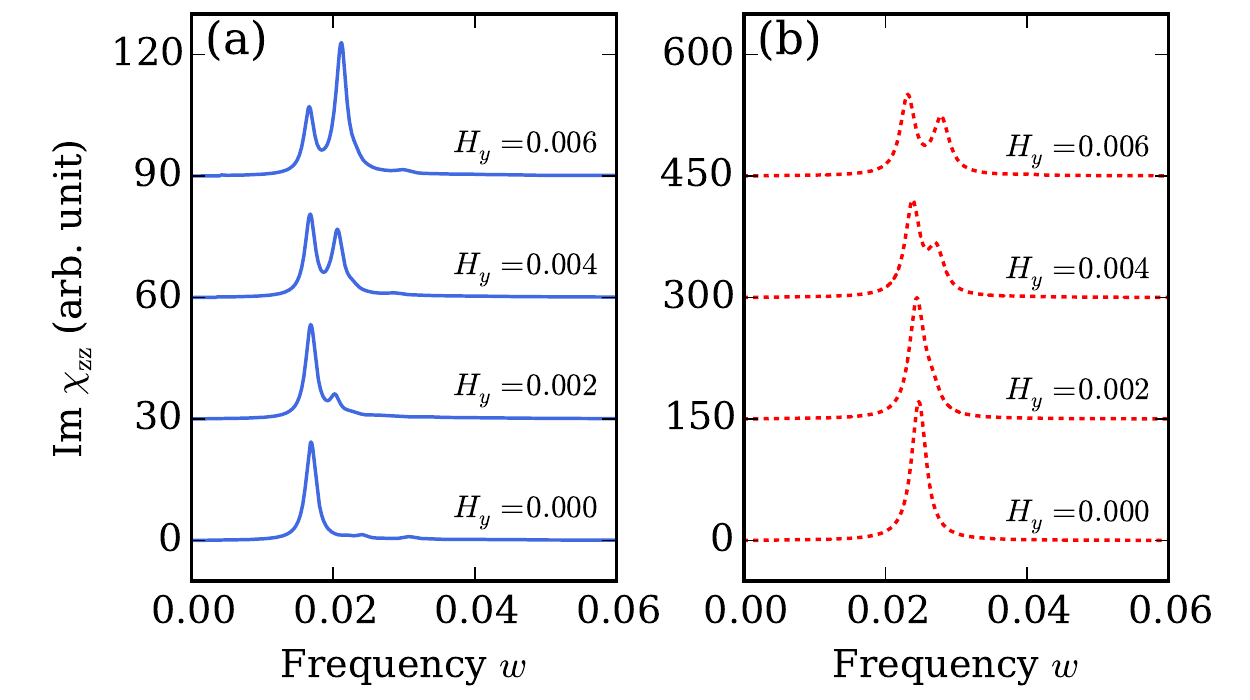}
\caption{Imaginary parts of the $H_y$-dependent dynamical susceptibility $\chi_{zz}$ as a function of frequency
for (a) a single skyrmion, and (b) a skyrmion lattice. The spectra are obtained by applying a sinc-field pulse $h = h_0 \mathrm{sinc}(\omega_0 t)$ to the system
with $h_0=1\times10^{-5}$ and $\omega_0=0.1 \pi$ in the $z$-axis, the magnetization dynamics is recorded every $dt=5$ for 8000 steps.
}
\label{fig_spectra}
\end{center}
\end{figure}

A two-dimensional system of size $N=160 \times 160$ sites with periodic boundary conditions is selected to study the
dynamics of a single skyrmion, Fig.~\ref{fig_skx}(a), and $N = 174 \times 150$ sites for the
skyrmion lattice, as shown in Fig.~\ref{fig_skx}(d).
We have chosen $J=\hbar=\gamma=S=a=1$ as simulation parameters~\cite{Mochizuki2012, Zhang2015}, therefore, the coefficients to convert the
external field $H$, time $t$, frequency $\omega$ and velocity $v$ to
SI units are $\unit{H} = J/\hbar \gamma S$,  $\unit{t} = \hbar S/J$,
$\unit{\omega} = J/\hbar S$ and $\unit{v} = J a/\hbar S$, respectively.
Table~I shows the expressions, and particular values for the case of $J=1$ meV, $S=1$ and
$a=0.5$ nm. We use simulation units throughout the paper. The perpendicular component $H_z$ is 
fixed as $H_z=0.02$ which corresponds to $0.173$ T for $S=1$ and $J=1$ meV.
We use Gilbert damping $\alpha=0.02$ for all simulations except for the magnetic spectra shown in Fig.~\ref{fig_spectra} where
$\alpha=0.04$ is chosen.
For the single skyrmion dynamics, we apply the absorbing boundary conditions for damping~\cite{Wang2015}
by setting $\alpha=1.0$ for the $20$ spins at the edges of a simulated domain.

The configuration of a skyrmion in the presence of an in-plane field $H_y=0.006$ is shown in
Fig.~\ref{fig_skx}(a). It is found that the radial symmetry is broken. Indeed, as shown in Fig \ref{fig_skx}(c),
the corresponding distribution of the topological charge density $q(x,y)=(1/4\pi)\mvec{m} \cdot (\partial_x \mvec{m} \times \partial_y \mvec{m})$ is asymmetric.
However, the total topological charge of a single skyrmion remains constant $Q=\int q dx dy=-1$. As a comparison,
Fig.~\ref{fig_skx}(b) shows the topological charge density $q$ for a skyrmion with radial symmetry when $H_y=0$.
Similar to the vortex~\cite{Cowburn1999}, the distortion of the skyrmion is along the $x$-axis when an external field is applied in the $y$-direction.

\begin{table}
\caption{\label{tab1} Unit conversion table for $J=1$ meV, $S=1$ and $a=0.5$ nm.}
\begin{ruledtabular}
\begin{tabular}{lll}
Distance $x$ & $\unit{x} = a$ & $ = 0.5$ nm\\
Time $t$ & \unit{t} = $\hbar S/J$ & $\approx 0.66$  ps \\
Velocity $v$ & $\unit{v} = Ja/(hS)$ & $\approx 7.59 \times 10^{2}$ m/s\\
Frequency $\omega$ & $\unit{\omega} = J/(\hbar S)$  & $\approx 1.52 \times 10^{3}$ GHz \\
Magnetic field $H$ &$\unit{H} = J/(\hbar \gamma S)$ & $\approx 8.63$ T \\
\end{tabular}
\end{ruledtabular}
\end{table}

The excitation of internal modes depends on the static external field $H_z$ as well as the frequency and
direction of microwaves~\cite{Lin2014a,Mochizuki2012}.
The typical excited modes are the clockwise/counterclockwise rotation and breathing modes~\cite{Mochizuki2012, Onose2012, Okamura2013}.
To study how the in-plane applied field $H_y$ affects the excitation mode of a skyrmion, we calculate the magnetic absorption spectrum of the skyrmion.
After applying a sinc-function field pulse $h = h_0 \mathrm{sinc}(\omega_0 t) = h_0 \sin(\omega_0 t)/(\omega_0 t)$ to the stable skyrmion state
we record the spatially averaged magnetization evolution and from that we compute the dynamic susceptibility $\chi$
via a Fourier transformation~\cite{Liu2008, Kim2014}.
For instance, the component $\chi_{zz}$ is computed using $m_z$ when the pulse is parallel to the $z$-axis.

Figure~\ref{fig_spectra} shows the imaginary part of the
dynamical susceptibility $\chi_{zz}$ for a single skyrmion (a) and a skyrmion
lattice (b); each calculated for different in-plane fields
$H_y$. We see in (b) for the skyrmion lattice and $H_y=0$ that the mode
with frequency $\omega \approx 0.0246$ is dominant. This is the so-called
breathing mode~\cite{Onose2012, Mochizuki2012}.
The resonance angular frequency $\omega \approx 0.0246$ in simulation
units corresponds to the frequency
$f=\omega/2 \pi \approx 5.95$~GHz (using $\unit{\omega}$ from Table~\ref{tab1}). The breathing mode frequency decreases slightly as the in-plane field $H_y$ increases.
A second peak emerges with increasing $H_y$, and the frequency of the new mode is $\omega \approx 0.0278$ when $H_y=0.006$.
Similar to the skyrmion lattice, a breathing mode with $\omega \approx
0.0168$ is found for a single skyrmion, as shown in
Fig.~\ref{fig_spectra}(a). In general, the resonance frequency of a
single skyrmion is lower than that for the skyrmion lattice.
As for the skyrmion lattice, a new mode with frequency around $0.02$
emerges for the single skyrmion case as $H_y$ is increased. This
second mode is the uniform mode with frequency $\omega=\gamma H$ where $H\simeq(H_y^2+H_z^2)^{1/2}$ is the amplitude of the external field.

\begin{figure}[tbhp]
\begin{center}
\includegraphics[scale=0.65]{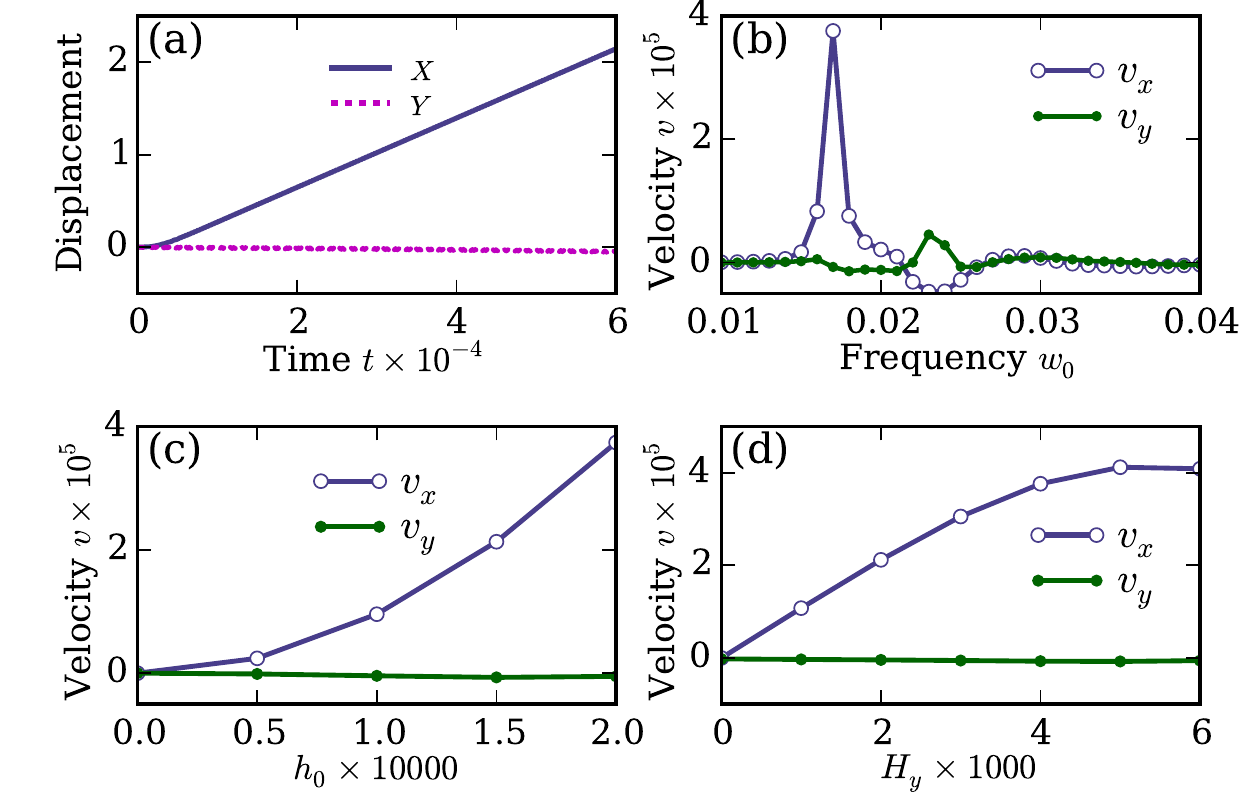}
\caption{ (a) The displacements of the guiding center $(X,Y)$ for a single skyrmion.
The simulation parameters are $\omega_0=0.017$, $h_0=2\times 10^{-4}$ and the in-plane field $H_y=0.004$.
(b-d) The velocities $v_x$ and $v_y$ as functions of (b) the microwave frequency $\omega_0$, (c) the microwave amplitude $h_0$,
and (d) the in-plane field $H_y$.
The fixed simulation parameters are the same as in (a).
}
\label{fig_dis}
\end{center}
\end{figure}

In the presence of an in-plane applied field $H_y$, the skyrmion is deformed, as shown in Figs.~\ref{fig_skx}(a) and \ref{fig_skx}(c). Therefore,
instead of the geometric center we measure the so-called guiding center~\cite{Papanicolaou1991} $\mvec{R}=(X,Y)$ of a skyrmion:
$X = \int x q d x d y/\int q d x d y$ and
$Y = \int y q d x d y/\int q d x d y$,
where $q$ is the topological charge density.
For a symmetric skyrmion, the guiding center is the same as its geometric center. 
In the rest of this work, we consider the scenario that a linearly polarized microwave is applied in the $z$-direction, 
i.e., $\mvec{h}(t)= h_0 \sin(\omega_0 t) \mvec{e}_z$, where $h_0$ and $\omega_0$ are the amplitude
and frequency of the microwave, respectively.

Figure~\ref{fig_dis}(a) shows the displacement of the guiding center for a single skyrmion with $\omega_0=0.017$ and
in-plane field $H_y=0.004$. The microwave amplitude is $h_0=2\times 10^{-4}$, which corresponds to
1.73 mT for $J=1$ meV and $S=1$. It can be seen that the $x$-component of the guiding center, $X$,
changes significantly as a function of time, while
the displacement of $Y$ is relatively small. 
Figure~\ref{fig_dis}(b) plots the frequency-dependent
skyrmion velocity and shows that a single skyrmion has maximum velocity when $\omega_0=0.017$,
which is the breathing mode resonance frequency. Therefore, exciting the breathing mode can move the skyrmion effectively
in the presence of an in-plane field $H_y$.
Using the conversions presented in Table~I, the maximum velocity is $v_x \approx 2.8$~cm/s.
While $v_x$ is positive for frequency $\omega_0 = 0.017$, it is negative for $\omega_0 = 0.023$,
where the former corresponds to the breathing mode, and the latter to coherent rotation.
The relation between the velocities and the amplitude of the microwave is shown in Fig.~\ref{fig_dis}(c),
and the dependence of $v_x$ on $h_0$ is nonlinear: $v_x \propto h_0^2$ which is proportional to the power of the microwaves.

Figure~\ref{fig_dis}(d) describes the relation between the skyrmion velocity and the in-plane field $H_y$.
The velocity is zero if $H_y=0$, which is expected due to the symmetry of the skyrmion.
The velocity of the skyrmion also depends on the direction of the in-plane field $H_y$:
 the velocity is reversed when the direction of the in-plane field is reversed.
Similarly, a change in the sign of the DMI constant will also reverse the sign of the velocity,
which is different from the case of driving skyrmions with spin-polarized currents, 
where the sign of perpendicular velocity (with respect to the current direction) of
the skyrmion motion is related to the sign of topological charge rather than the DMI constant sign.

\begin{figure}[tbhp]
\begin{center}
\includegraphics[scale=0.6]{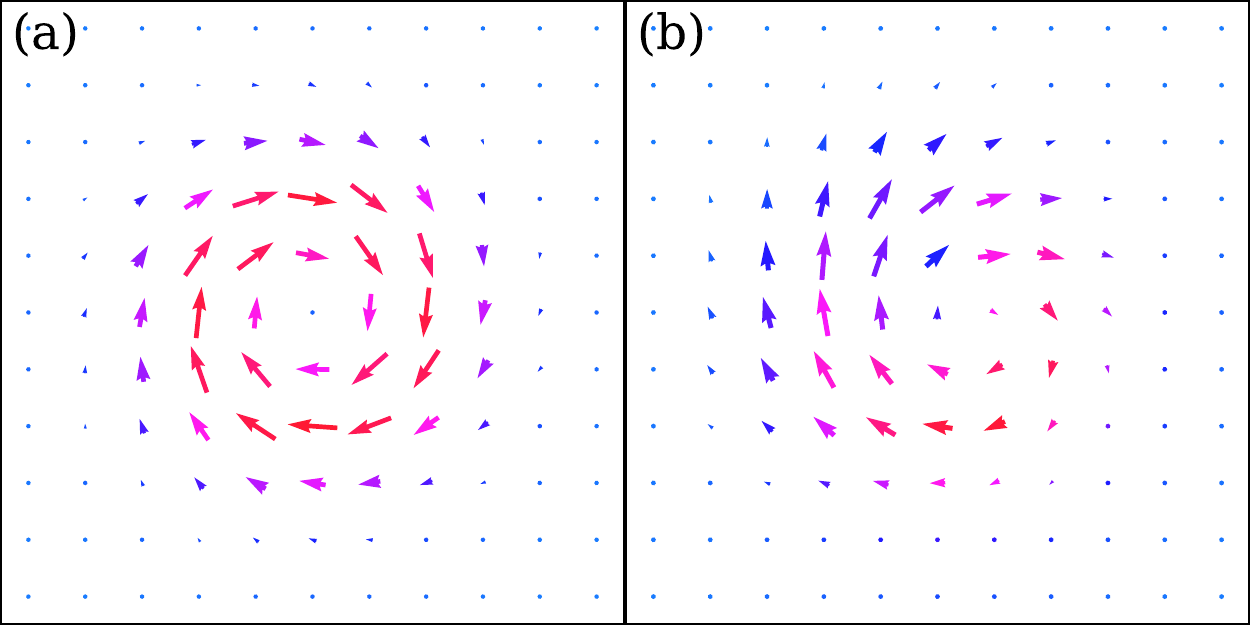}
\caption{The total spatial force density $f_i=\tilde{\mvec{m}}_s \cdot [ \partial_i \tilde{\mvec{m}}_s \times \mean{\mvec{m} \times \Heff}]$ 
for (a) $H_y=0$, and (b) $H_y=4 \times 10^{-3}$, where we have used $\tilde{\mvec{m}}_s=\mean{\mvec{m}}$.
 The microwave frequency is $\omega_0=0.017$.  }
\label{fig_sc}
\end{center}
\end{figure}

To understand why the skyrmion moves in the presence of an in-plane field, we split the magnetization
unit vector $\mvec{m}$ into a slow part $\mvec{m}_s$ and a fast part $\mvec{n}$, i.e., $\mvec{m} = \mvec{m}_s + \mvec{n}$,  
where the slow part represents the equilibrium profile of the skyrmion while the fast part is 
responsible for the excited spin wave mode~\cite{Kong2013, Kovalev2014}.
In the continuum approximation, the effective field is $\Heff= \tilde{A} \nabla^2 \mvec{m} - \tilde{D} \nabla \times \mvec{m} + \mvec{H} +\mvec{h}(t)$,
where $\tilde{A}=2A/M_s$ and $\tilde{D}=2D_a/M_s$. In the presence of the microwaves with specific frequency, bound spin-wave modes
are excited, and thus we expect $\mean{\dot{\mvec{n}}}=0$ and $\mean{\mvec{n}}=0$ due to the microwave synchronization, where $T=2\pi/\omega_0$ is
the period of microwaves and the notation $\mean{f}=\mean{f}(t) \equiv T^{-1}\cdot \int_t^{t+T} \! f(t') d t'$ represents 
the time average of function $f(t)$ over a single period $T$. 
Furthermore, one obtains $\mean{\mvec{m}_s \times \dot{\mvec{n}}} \approx 0$ and $\mean{\mvec{n} \times \dot{\mvec{m}}_s} \approx 0$ since $\mvec{m}_s$ 
is the slow part. Therefore, by averaging the LLG equation (\ref{eq_llg}) over a period $T$ we arrive at
\begin{equation}\label{eq_llg2}
\mean{ \dot{\mvec{m}}_s}= -\gamma \mean{ \mvec{m} \times \Heff} + \alpha \mean{\mvec{m}_s \times \dot{\mvec{m}}_s},
\end{equation}
where $\mean{\mvec{n} \times \dot{\mvec{n}}}=0$ is used since basically $\mvec{n}$ is a sine or cosine function in time. 
We then consider the possible translational motion of the skyrmion such that $\mvec{m}_s(\mvec{r},t)=\mvec{m}_s(\mvec{r}-\mvec{v}_s t)$,
i.e., $\dot{\mvec{m}}_s=-(\mvec{v}_s \cdot  \nabla) \mvec{m}_s$, where the skyrmion velocity $\mvec{v}_s=d \mvec{R}/ dt$ 
is assumed to be a constant.
If the skyrmion moves slowly, i.e., $v_s T \ll L$ ($L$ is the typical skyrmion size), we have 
$ \mean{\dot{\mvec{m}}_s} \approx -(\mvec{v}_s \cdot  \nabla) \tilde{\mvec{m}}_s$ (see Appendix \ref{appendix_A}) 
where $\tilde{\mvec{m}}_s =\mean{\mvec{m}_s}$. 
Similarly, $\mean{\mvec{m}_s \times \dot{\mvec{m}}_s} \approx \tilde{\mvec{m}}_s \times \mean{\dot{\mvec{m}}_s}$, and thus
Eq.~(\ref{eq_llg2}) can be rewritten as 
\begin{equation}\label{eq_llg3}
 (\mvec{v}_s \cdot  \nabla) \tilde{\mvec{m}}_s = \gamma \mean{ \mvec{m} \times \Heff} 
+ \alpha \tilde{\mvec{m}}_s \times (\mvec{v}_s \cdot  \nabla) \tilde{\mvec{m}}_s.
\end{equation}
Following Thiele's approach in describing the motion of magnetic textures~\cite{Thiele1973}, 
we replace the dots in $\int \tilde{\mvec{m}}_s \cdot (\partial_i \tilde{\mvec{m}}_s\times \cdots)dx dy$ 
by Eq.~(\ref{eq_llg3}) to obtain~\cite{Everschor2012, Kong2013, Kovalev2014}
\begin{equation}\label{eq_motion}
\mvec{G} \times  \mvec{v}_s  + \widehat{\mathcal{D}}  \mvec{v}_s  = \mvec{F},
\end{equation}
where $i=x,y$ and $\mvec{G}=4 \pi Q \mvec{e}_z$.
The tensor $\widehat{\mathcal{D}}_{ij}=\alpha \eta_{ij}$ is the damping tensor in which
$\eta_{ij}= \int (\partial_{i} \tilde{\mvec{m}}_s \cdot \partial_{j} \tilde{\mvec{m}}_s) dx dy = \delta_{ij} \eta$
is the shape factor of the skyrmion and $\eta$ is close to $4
\pi$~\cite{Kovalev2014}.
The force $\mvec{F}$ is given by
\begin{equation}\label{eq_force}
F_i = - \gamma \int \tilde{\mvec{m}}_s \cdot \big[ \partial_i \tilde{\mvec{m}}_s \times \mean{\mvec{m} \times \Heff} \big] dx  dy.
\end{equation}

Figure~\ref{fig_sc}(a) and (b) depict the total spatial force density for $H_y=0$ and $H_y=4 \times 10^{-3}$, respectively, 
where we have used $\tilde{\mvec{m}}_s=\mean{\mvec{m}}$.
The force density is symmetric if $H_y=0$ and thus the total force $\mvec{F}$ is zero. 
However, when $H_y$ is nonzero the force distribution is asymmetric which results in the 
skyrmion motion due to the nonzero net force. For small damping $\alpha \ll 1$, we have $v_x \approx F_y/(4\pi Q)$.
The total force calculated with parameters $\omega_0=0.017$, $H_y=0.004$ and $h_0=2\times 10^{-4}$ is $F_y=-4.7\times 10^{-4}$,
therefore, the established velocity is $v_x=3.7\times 10^{-5}$,  which fits the simulation result ($\sim 3.8 \times 10^{-5}$) well.
Similarly, for $\omega = 0.023$ using Eq.~(\ref{eq_force}) we obtain $F_y = 5.7\times10^{-5}$ and find $v_x \approx -4.5\times10^{-6}$ from Eq.~(\ref{eq_motion}); 
in agreement with the simulation results (the minimum of $v_x$ is $-4.7 \times 10^{-6}$). 

It is of interest to circumstantiate the contributions of the total force $\mvec{F}$. 
In Appendix \ref{appendix_B} we show that there are three nontrivial terms
\begin{equation}\label{eq_heff}
\mean{\mvec{m} \times \Heff}  \approx \mean{\mvec{n} \times [\tilde{A} \nabla^2 \mvec{n} - \tilde{D} \nabla \times \mvec{n} 
+ \mvec{h}(t) ]}.
\end{equation}
The exchange term $\mvec{n} \times \tilde{A} \nabla^2 \mvec{n}$ corresponds to magnon currents~\cite{Kong2013, Kovalev2014}. 
Compared to the skyrmion motion induced by the temperature gradient, where the magnon current is
generated by the temperature gradient, here the magnon current originates from the external microwave fields.
Another difference is that in our case the contributions from DMI and microwave fields are also significant. 

\begin{figure}[tbhp]
\begin{center}
\includegraphics[scale=0.65]{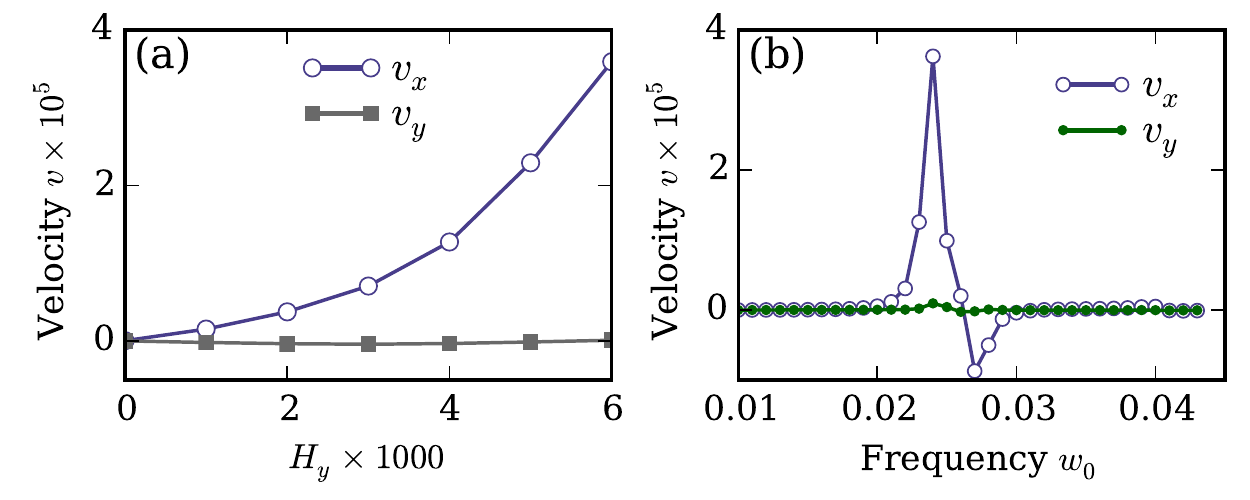}
\caption{
(a) The velocities $v_x$ and $v_y$ as a function of an in-plane field $H_y$ for the skyrmion lattice at frequency $\omega_0=0.023$.
(b) The velocities as a function of frequency $\omega_0$ with $H_y=4 \times 10^{-3}$.
The microwave amplitude is $h_0=2\times 10^{-4}$.
}
\label{fig_v_lattice}
\end{center}
\end{figure}

We repeat the velocity study for the skyrmion lattice.
Fig.~\ref{fig_v_lattice}(a) plots the velocities $v_x$ and $v_y$ of the skyrmion lattice as functions
of the in-plane external field $H_y$. The dependencies are similar to the single skyrmion case.
The frequency-dependent velocities $v_x$ and $v_y$ are shown in Fig.~\ref{fig_v_lattice}(b).
As for the single skyrmion case, the velocity peak coincides with the dominant dynamical
susceptibility peak in Fig.~\ref{fig_spectra}(b).

In closing, we briefly comment on the importance of symmetry breaking in driving the skyrmions. 
The driving force originates from the microwave field, which is periodic in time and averages to zero. 
The symmetry-breaking field converts the periodic microwave field into a net force and thus moves skyrmions effectively.
A related field with periodic driving forces is that of ratchet-like transport phenomena \cite{Reimann2002, Hanggi2009, Quintero2010}, 
where the net motion is obtained by breaking the spatial symmetry \cite{Reimann2002} or temporal symmetry \cite{Quintero2010}. 
We also note that preliminary simulation results suggest that for magnon-driven skyrmions 
\cite{Iwasaki2014, Schutte2014} the introduction of a symmetry-breaking in-plane field affects 
the skyrmion's motion and changes the Hall angle significantly.

In summary, we have studied the skyrmion dynamics driven by microwaves in the presence of an in-plane external field.
We found that both a single skyrmion and a skyrmion lattice  can be moved by a linearly polarized microwave field if the 
axial symmetry of skyrmions is slightly broken. These results suggest a novel method for skyrmion manipulation using microwaves fields.

We acknowledge financial support from EPSRC's DTC grant EP/G03690X/1.
W.W. thanks the China Scholarship Council for financial support.
The authors acknowledge the use of the IRIDIS High Performance Computing Facility, and associated support services at the University of Southampton, 
in the completion of this work.

\appendix
\section{}
\label{appendix_A}
Assume that a well-behaved function $f(x,t)=f(x-vt)$ describes the dynamics of a soliton where $v$ is a constant.
As we can see, $f$ satisfies $\dot{f}=-vf'$. 
For given time $T$, if $v T \ll L$ where $L$ is the typical size of the soliton (for example, $L$ could be the 
domain wall width for a magnetic domain wall), we can find that
\begin{eqnarray}
\mean{\dot{f}}(0) = \frac{1}{T} \int_0^T \! \dot{f} dt= \frac{1}{T}[f(x-vT)-f(x)] \nonumber \\
\approx -v f'(x-vT/2),
\end{eqnarray}
where we have used the Taylor series for $f(x)$ and $f(x-vT)$:
\begin{eqnarray}
&f(x) \approx f(x-vT/2) + f'(x-vT/2) vT/2 \\
&f(x-vT) \approx f(x-vT/2) - f'(x-vT/2) vT/2.
\end{eqnarray}
Similarly, we can see that $\tilde{f}_0 \equiv \mean{f}(0) \approx f(x-vT/2)$ and thus we have $\mean{\dot{f}}(0) \approx -v \tilde{f}_0$.
This relation actually holds for arbitrary $t$
\begin{equation}
\mean{\dot{f}} \approx -v \tilde{f}.
\end{equation}

\section{}
\label{appendix_B}
By using the effective field explicitly and noticing that $\mvec{m}=\mvec{m}_s+\mvec{n}$, 
the term $\mean{\mvec{m} \times \Heff}$ can be splited into four parts
\begin{equation}
\mean{\mvec{m} \times \Heff} =  \mean{T_1 + T_2 + T_3 + T_4},
\end{equation}
where $T_1=\mvec{n} \times [\tilde{A} \nabla^2 \mvec{n} - \tilde{D} \nabla \times \mvec{n} + \mvec{h}(t) ]$ 
is shown in Eq~(\ref{eq_heff}), $T_2= \mvec{m}_s \times (\tilde{A} \nabla^2 \mvec{m}_s - \tilde{D} \nabla \times \mvec{m}_s + \mvec{H})$,
$T_3= \mvec{n} \times (\tilde{A} \nabla^2 \mvec{m}_s - \tilde{D} \nabla \times \mvec{m}_s + \mvec{H})$ and 
$T_4=\mvec{m}_s \times [\tilde{A} \nabla^2 \mvec{n} - \tilde{D} \nabla \times \mvec{n} + \mvec{h}(t) ]$.
We expect $T_2=0$ since $\mvec{m}_s$ represents the equilibrium state of the skyrmion. 
For the slow skyrmion motion, replacing $\mvec{m}_s$ by $\tilde{\mvec{m}}_s$ and noticing 
that $\mean{\mvec{n}}=0$, we obtain $\mean{T_3}\approx \mean{\mvec{n}} \times (\tilde{A} \nabla^2 \tilde{\mvec{m}}_s - \tilde{D} \nabla \times \tilde{\mvec{m}}_s)=0$ 
and $\mean{T_4} \approx \tilde{\mvec{m}}_s \times \mean{\tilde{A} \nabla^2 \mvec{n} - \tilde{D} \nabla \times \mvec{n} + \mvec{h}(t)}=0$.
In this slow motion approxmation, the fast part $\mvec{n}$ can be computed as $\mvec{n}\approx \mvec{m}-\mean{\mvec{m}}$.

\bibliographystyle{apsrev4-1}
%

\end{document}